\documentclass[10pt,conference,letterpaper]{IEEEtran}
\usepackage[letterpaper,left=0.65in,right=0.65in,top=0.65in,bottom=0.75in]{geometry}

\usepackage{amsmath,amssymb}
\usepackage{tikz-cd}
\usepackage{multicol}
\usepackage{url}
\usepackage{array}
\usepackage{graphicx}

\title{\Large SCALE-Sim EVA: Design Principles for an Extensible, Visualizable, and Adaptable Accelerator Simulation Framework}
\author{
\IEEEauthorblockN{Jingtian Dang, Ritik Raj, Tushar Krishna}
\IEEEauthorblockA{
Georgia Institute of Technology\\
dangjingtian@gatech.edu, ritik.raj@gatech.edu, tushar@ece.gatech.edu
}
}
\begin{document}
\maketitle
\vspace{-0.8em}


\section{Overview}

Modern AI and HPC systems increasingly rely on heterogeneous accelerator architectures that combine systolic arrays, vector units, hierarchical memories, local buffers, and specialized data movement paths. Modeling these systems requires reasoning about compute latency, memory service time, tensor placement, data readiness, resource availability, and compiler-driven mapping choices. Existing accelerator simulators provide useful cycle-level insight, but many are tied to specific execution models, hardware organizations, or workload abstractions.

This abstract presents SCALE-Sim EVA, an extensible, visualizable, and adaptable framework for IR-aware accelerator simulation. EVA uses a lightweight abstraction stack for accelerator ModSim: commands describe scheduled tensor operations, tensor handles track runtime storage state, hardware components execute commands through configurable handlers, and traces expose command and storage behavior over time.

\section{EVA Framework and Timing Model}

\begin{figure}[!t]
    \centering
    \includegraphics[width=\columnwidth]{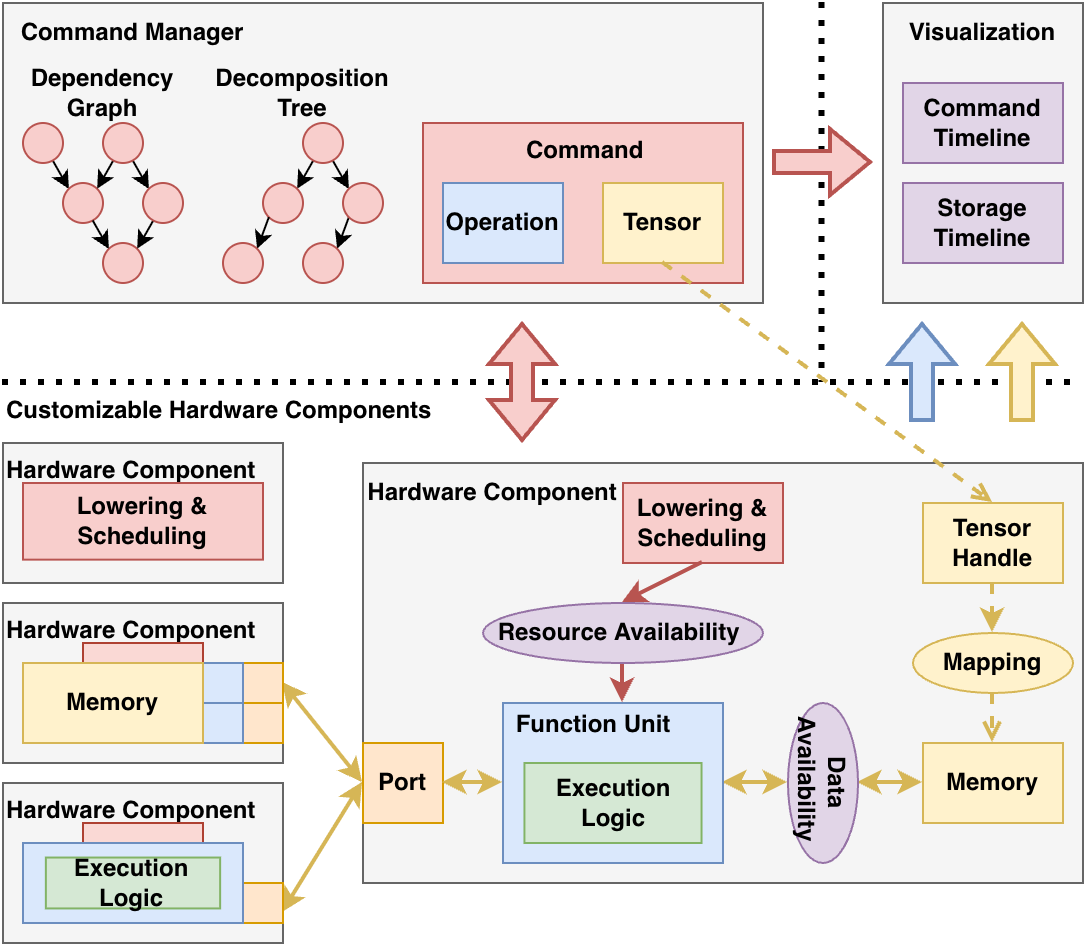}
    \caption{Overview of the EVA framework. EVA separates command management, customizable hardware components, tensor/storage state, and visualization. Commands execute when both input data and target hardware resources are available, and EVA records command and storage timelines for analysis.}
    \label{fig:eva_overview}
\end{figure}

Figure~\ref{fig:eva_overview} shows the overall organization of EVA. The command manager maintains dependency graphs and decomposition trees over commands. Each command contains an operation and the tensors involved in that operation, but does not prescribe a fixed execution granularity. A hardware component may execute a command directly using a matching function unit, or decompose it into smaller child commands spanning multiple components. This allows the same framework to support different tensor IR levels, from coarse-grained operators such as attention to fine-grained data movement and elementwise commands.

EVA does not assume a fixed accelerator organization. Instead, users can define hardware components with different internal capabilities and connect them through ports to form a user-defined hardware topology.

The key timing principle is the interaction between data availability and hardware resource availability. A command can begin only when its input tensor data is ready and its target hardware resource can accept new work. Its completion then updates downstream tensor readiness and hardware availability. This explicit timing state preserves cycle-level reasoning without hard-coding the simulator around one accelerator organization. EVA records command execution and tensor storage behavior to generate command timelines and storage timelines for visualization.

\section{Related Work}

\begin{table*}[t]
\centering
\caption{Positioning of EVA relative to representative modeling/simulation frameworks.}
\label{tab:comparison}
\small
\renewcommand{\arraystretch}{1.15}
\begin{tabular}{>{\raggedright\arraybackslash}m{0.16\textwidth} >{\raggedright\arraybackslash}m{0.25\textwidth} >{\raggedright\arraybackslash}m{0.25\textwidth} >{\raggedright\arraybackslash}m{0.25\textwidth}}
\hline
\textbf{Framework} &
\textbf{Target / style} &
\textbf{Main abstraction} &
\textbf{Relation to EVA} \\
\hline
SCALE-Sim v3~\cite{raj2025scale} &
Modular cycle-level simulation for end-to-end systolic accelerator systems &
Systolic arrays, dataflows, memory hierarchy, and system modules &
EVA generalizes cycle-level reasoning beyond systolic-array-centered execution using commands, tensors, and hardware services. \\
\hline
Timeloop~\cite{timeloop} &
Analytical modeling and mapping search for tensor-algebra accelerators &
Loop nests, tiling, and spatial/temporal mappings &
EVA focuses on command-level execution, runtime tensor state, and data/resource readiness after a mapping is chosen. \\
\hline
Accel-Sim~\cite{accelsim} &
Trace-driven simulation for GPU-like programmable accelerators &
GPU kernels, warps, SMs, and memory hierarchy &
EVA is tensor/IR-centric rather than GPU thread/warp-centric. \\
\hline
gem5-SALAM~\cite{gem5salam} &
Full-system and microarchitectural simulation for heterogeneous systems &
CPUs, caches, OS/software stack, accelerators, scratchpads, and DMAs &
EVA is lighter-weight and targets tensor-level accelerator ModSim instead of full-system/ISA-level detail. \\
\hline
EVA (this work) &
Command-driven cycle-level design for AI/HPC accelerator ModSim &
Logical tensors, tensor handles, tensor mappings, command trees, hardware services, and traces &
Provides an extensible, visualizable, and adaptable abstraction stack for IR-aware accelerator modeling. \\
\hline
\end{tabular}
\end{table*}

Table~\ref{tab:comparison} positions EVA relative to these frameworks. EVA complements them by targeting IR-aware, tensor-centric cycle-level ModSim after a mapping is chosen, while remaining lighter than full-system simulators and more general than systolic-array-centered designs.

\section{Customizable Hardware Components}

EVA models hardware components as customizable modules that combine function units, memory modules, and lowering and scheduling logic. Function units define command-execution behavior, memory modules define tensor storage behavior, and lowering and scheduling logic determines how high-level commands are decomposed, mapped, and ordered.

A component may contain multiple function units and memory modules, allowing users to model heterogeneous compute engines, local buffers, scratchpads, caches, or specialized data movement paths within the same hardware component. Larger accelerator systems can be built by connecting user-defined components through ports that carry tensor data and commands. This keeps the hardware organization flexible: users can compose systolic-array components, vector components, memory hierarchies, and specialized accelerators without changing the core EVA command and tensor abstractions.

Components can also communicate through messages. Users may define both message types and handlers, allowing custom inter-component protocols.

\section{Decomposable IR Execution}

A key design goal of EVA is to support commands at different tensor IR granularities. A command may represent fine-grained data movement, elementwise work, or coarse-grained tensor operators such as matrix multiplication, convolution, or attention without changing the core simulation interface.

Decomposition exposes parallelism, memory movement, and heterogeneous hardware utilization across a command tree. A coarse command enables fast architectural exploration when a specialized unit is available, while a decomposed command tree reveals tiling, scheduling, memory traffic, and resource contention. This flexibility bridges compiler-visible tensor IRs and hardware-specific execution models.

\section{Tensor and Memory Abstractions}

EVA separates tensor semantics from physical storage. Logical tensors record shape, datatype, and identity, while runtime tensor handles track location, readiness, and active mappings. Tensor mappings express views such as slicing, tiling, transposition, broadcast, or padding without requiring each derived tile to own a separate physical allocation.

The memory abstraction resolves mapped tensor indices into hardware addresses through memory regions and layouts. This keeps tile writeback deterministic while allowing row-major, blocked, padded, sharded, sparse, or compressed layouts without changing the core tensor abstraction.

\section{Why This Design Matters}

EVA's design matters along three axes. It is \emph{extensible}: new tensor operations, hardware components, memory behaviors, and mapping policies can be expressed through the same command, tensor, and component abstractions. It is \emph{visualizable}: command and storage traces expose when work executes, where tensor data resides, and how data readiness and resource availability shape execution over time.

It is also \emph{adaptable}: the same framework can model coarse tensor operators, decomposed command trees, and different hardware organizations by changing component behavior rather than changing the overall timing model. This is especially useful for IR-aware simulation, where different compiler mappings can generate different command trees over the same hardware model for design-space exploration.

\section{Conclusion}

SCALE-Sim EVA presents a design and methodology contribution for IR-aware accelerator ModSim. Its command-driven, tensor-aware, and component-based structure provides a modular path beyond fixed accelerator execution models.

Its expected impact is to make it easier to connect compiler mappings, hardware models, memory hierarchy studies, and visual debugging within one extensible simulation framework.

\bibliographystyle{IEEEtran}
\bibliography{references}

\end{document}